\documentclass[preprint2]{aastex6}

\usepackage{amsmath,amssymb,natbib}
\usepackage{microtype}
\usepackage{hyperref}
\usepackage{graphicx}
\usepackage[caption=false]{subfig}
\newcommand{\sbunit}{mag~arcsec$^{-2}$}
\newcommand{\HI}{H{\sc\,i}}

\newcommand{\LCDM}{$\Lambda$CDM}
\newcommand{\OIII}{[O\,{\sc iii}]$\,\lambda 5007$}
\newcommand{\NIIa}{[N\,{\sc ii}]$\,\lambda 6549$}
\newcommand{\NIIb}{[N\,{\sc ii}]$\,\lambda 6583$}
\newcommand{\NII}{[N\,{\sc ii}]$\,\lambda\lambda6549,6583$}

\newcommand{\code}[1]{\texttt{#1}}

\newcommand{\resp}[1]{#1}

\newcommand{\m}[1]{\mathrm{#1}}

\definecolor{citecolor}{rgb}{0,0.1,0.43}
\definecolor{linkcolor}{rgb}{0,0.1,0.43}
\hypersetup{backref,breaklinks,colorlinks,
urlcolor=citecolor,linkcolor=linkcolor,citecolor=citecolor}

\graphicspath{{figures/}}
 
\shorttitle{Two Diffuse Dwarf Galaxies in the Field}
\shortauthors{Greco et al.}

\begin{document}\sloppy\sloppypar\raggedbottom\frenchspacing 

\title{A Study of Two Diffuse Dwarf Galaxies in the Field}
\author{Johnny~P.~Greco\altaffilmark{\pu},
        Andy~D.~Goulding\altaffilmark{\pu},
        Jenny~E.~Greene\altaffilmark{\pu},
        Michael~A.~Strauss\altaffilmark{\pu},
        Song Huang\altaffilmark{\santa},
        Ji Hoon Kim\altaffilmark{\subaru},
        Yutaka Komiyama\altaffilmark{\naoj,\sokendai}
}
\email{jgreco@astro.princeton.edu}

% Affiliations
\newcommand{\pu}{1}
\newcommand{\santa}{2}
\newcommand{\subaru}{3}
\newcommand{\naoj}{4}
\newcommand{\sokendai}{5}
\altaffiltext{\pu}{Department of Astrophysical Sciences, 
Princeton University, Princeton, NJ 08544, USA}
\altaffiltext{\santa}{Department of Astronomy and Astrophysics, 
University of California, Santa Cruz, 
1156 High Street, Santa Cruz, CA 95064 USA}
\altaffiltext{\subaru}{Subaru Telescope, National Astronomical 
Observatory of Japan, 650 N Aohoku Pl, Hilo, HI 96720}
\altaffiltext{\naoj}{National Astronomical Observatory of Japan, 
2-21-1 Osawa, Mitaka, Tokyo 181-8588, Japan}
\altaffiltext{\sokendai}{Department of Astronomy, School of Science, 
Graduate University for Advanced Studies (SOKENDAI), 2-21-1, Osawa, Mitaka, 
Tokyo 181-8588, Japan}

\begin{abstract}
We present optical long-slit spectroscopy and far-ultraviolet to mid-infrared spectral energy distribution fitting of two diffuse dwarf galaxies, LSBG-285 and LSBG-750, which were recently discovered by the Hyper Suprime-Cam Subaru Strategic Program (HSC-SSP). We measure redshifts using H$\alpha$ line emission, and find that these galaxies are at comoving distances of ${\approx}25$ and ${\approx}41$~Mpc, respectively, after correcting for the local velocity field. They have effective radii of $r_\mathrm{eff}=1.2$ and 1.8~kpc and stellar masses of $M_\star\approx2$-$3\times10^{7}~M_\odot$. There are no massive galaxies ($M_\star>10^{10}~M_\odot$) within a comoving separation of at least 1.5~Mpc from LSBG-285 and 2~Mpc from LSBG-750. These sources are similar in size and surface brightness to ultra-diffuse galaxies, except they are isolated, star-forming objects that were optically selected in an environmentally blind survey. Both galaxies likely have low stellar metallicities $[Z_\star/Z_\odot] < -1.0$ and are consistent with the stellar mass--metallicity relation for dwarf galaxies. We set an upper limit on LSBG-750's rotational velocity of ${\sim}50$~km s$^{-1}$, which is comparable to dwarf galaxies of similar stellar mass with estimated halo masses $<10^{11}~M_\odot$. We find tentative evidence that the gas-phase metallicities in \resp{both of} these diffuse systems are high for their stellar mass, though a statistically complete, optically-selected galaxy sample at very low surface brightness will be necessary to place these results into context with the higher-surface-brightness galaxy population. 
\end{abstract}


\section{Introduction}

Low-surface-brightness (LSB) dwarf ($M_\star\lesssim10^9~M_\odot$) galaxies offer unique testing grounds for theories of galaxy formation and evolution. The relatively shallow gravitational potential wells of these systems make them highly sensitive to stellar feedback processes \citep[e.g.,][]{Larson:1974aa, Dekel:1986aa, Governato:2010aa, El-Badry:2016aa}, and as dark-matter dominated systems, their ultra-low stellar densities allow their dark matter distributions to be studied with little ambiguity from the challenges of quantifying the baryon component \citep[e.g.,][]{deBlok:2001aa, Marchesini:2002aa}. Moreover, their number densities and distribution of physical properties provide some of the most stringent tests of the dark energy plus cold dark matter (\LCDM) paradigm \citep[e.g.,][]{Weinberg:2015aa, Bullock:2017aa}. Yet, our census of this important population, particularly beyond the Local Group, remains highly incomplete because optical surveys generally suffer from strong surface-brightness selection effects \citep[e.g.,][]{Disney:1976aa, Blanton:2005aa}. 

The severity of this problem was recently underscored by the discovery of an abundant population of physically large, ultra-LSB galaxies in the Coma cluster (ultra-diffuse galaxies or UDGs; \citealt{van-Dokkum:2015aa, Koda:2015aa}). These diffuse galaxies are characterized by red colors, smooth ellipsoidal morphologies, \resp{optical central surface brightnesses fainter than ${\sim}24$~\sbunit, and effective radii $r_\mathrm{eff}>1.5$~kpc. While such objects have been known to exist for decades \citep[e.g.,][]{Sandage:1984aa, Dalcanton:1997aa, Conselice:2003aa}, their  abundance in clusters was not fully appreciated.} UDGs are now understood to be common in dense galaxy environments \citep[e.g.,][]{van-der-Burg:2016aa, van-der-Burg:2017aa}. More recently, a population of gas-rich, blue UDGs with irregular morphologies has been uncovered in low-density environments \resp{\citep[e.g.,][]{Roman:2017aa, Roman:2017ab, Trujillo:2017aa, Bellazzini:2017aa, Leisman:2017aa, Greco:2018ab}}. 

Most known isolated UDGs have been discovered via neutral hydrogen gas (\HI) using the ALFALFA survey \citep{Giovanelli:2005aa, Leisman:2017aa}; these objects are therefore generally gas-rich and star-forming. Two interesting exceptions are the optically-discovered objects R-127-1 and M-161-1 \citep{Dalcanton:1997aa}. Both of these galaxies are very isolated, yet they have quiescent optical spectra and low \HI\ content \citep{Papastergis:2017aa}. That they are both quenched and isolated is quite surprising, as essentially all known quenched dwarfs exist near a massive ($M_\star\gtrsim10^{10}~M_\odot$) neighbor \citep{Geha:2012aa}. This highlights the need for complete optically-selected galaxy samples at low surface brightness, which complement \HI\ searches that are biased to the most gas rich systems. Indeed, a combination of deep optical and \HI\ LSB galaxy surveys will be required to properly address pressing small-scale problems in standard cosmology such as the potential ``too big to fail'' problem \citep{Boylan-Kolchin:2011aa}  in the field \citep{Papastergis:2015aa}.

With this motivation, we are carrying out a blind search for LSB galaxies with the Hyper Suprime-Cam Subaru Strategic Program \citep[HSC-SSP;][]{Aihara:2018aa}, an ongoing optical wide-field survey using the Hyper Suprime-Cam \citep{Miyazaki:2018aa, Komiyama:2018aa, Furusawa:2018aa, Kawanomoto:inprep} on the Subaru Telescope. In our initial search of the first 200~deg$^2$ of the survey \citep{Greco:2018ab}, we uncovered ${\sim}$800 LSB galaxies, spanning a wide range of galaxy colors and environments. Our survey is deep (5$\sigma$ point-source detection of $i \sim26$~mag), wide (1400 deg$^2$ upon survey completion), and based on stellar continuum rather than gas, making it sensitive to both quenched and star-forming LSB galaxies.

As we show in \citet{Greco:2018ab}, our sample is diverse, ranging from dwarf spheroidals and UDGs in nearby groups to gas-rich irregulars to giant LSB spirals. To place our sample within the cosmological context, distance information will be essential. Therefore, we are undertaking a systematic follow-up program to map out the spatial distribution and physical properties of these galaxies. In this work, we present the results of a pilot study in which we obtained optical spectra of two galaxies (LSBG-285 and LSBG-750) from our sample, both of which turn out to be isolated, physically large ($r_\mathrm{eff}>1$~kpc) star-forming LSB dwarfs. These are the first two objects for which we have obtained spectroscopy, and they demonstrate that the HSC-SSP is sensitive to diffuse dwarf galaxies well beyond the Local Volume. 

We organize this paper as follows. In Section~\ref{sec:target}, we describe our target selection for this pilot study. We present our spectroscopic data and analysis in Section~\ref{sec:spec-ana}, and our photometric data and analysis in Section~\ref{sec:phot-ana}. In Section~\ref{sec:galprops}, we present the physical and environmental properties of LSBG-285 and LSBG-750, and in Section~\ref{sec:discussion} we discuss these objects in the context of the general dwarf galaxy and UDG  populations. We conclude with a summary in Section~\ref{sec:summary}.

Throughout this work, we assume a standard cosmology with $H_0=70$ km~Mpc$^{-1}$, $\Omega_m=0.3$, and $\Omega_\Lambda=0.7$. All magnitudes presented in this paper use the AB system \citep{Oke:1983aa}. Unless stated otherwise, we correct for Galactic extinction using the $E(B-V)$ values from the dust map of \citet{Schlegel:1998aa} and the recalibration from \citet{Schlafly:2011aa}.

\begin{figure*}[ht!]
    \centering
    \includegraphics[width=\textwidth]{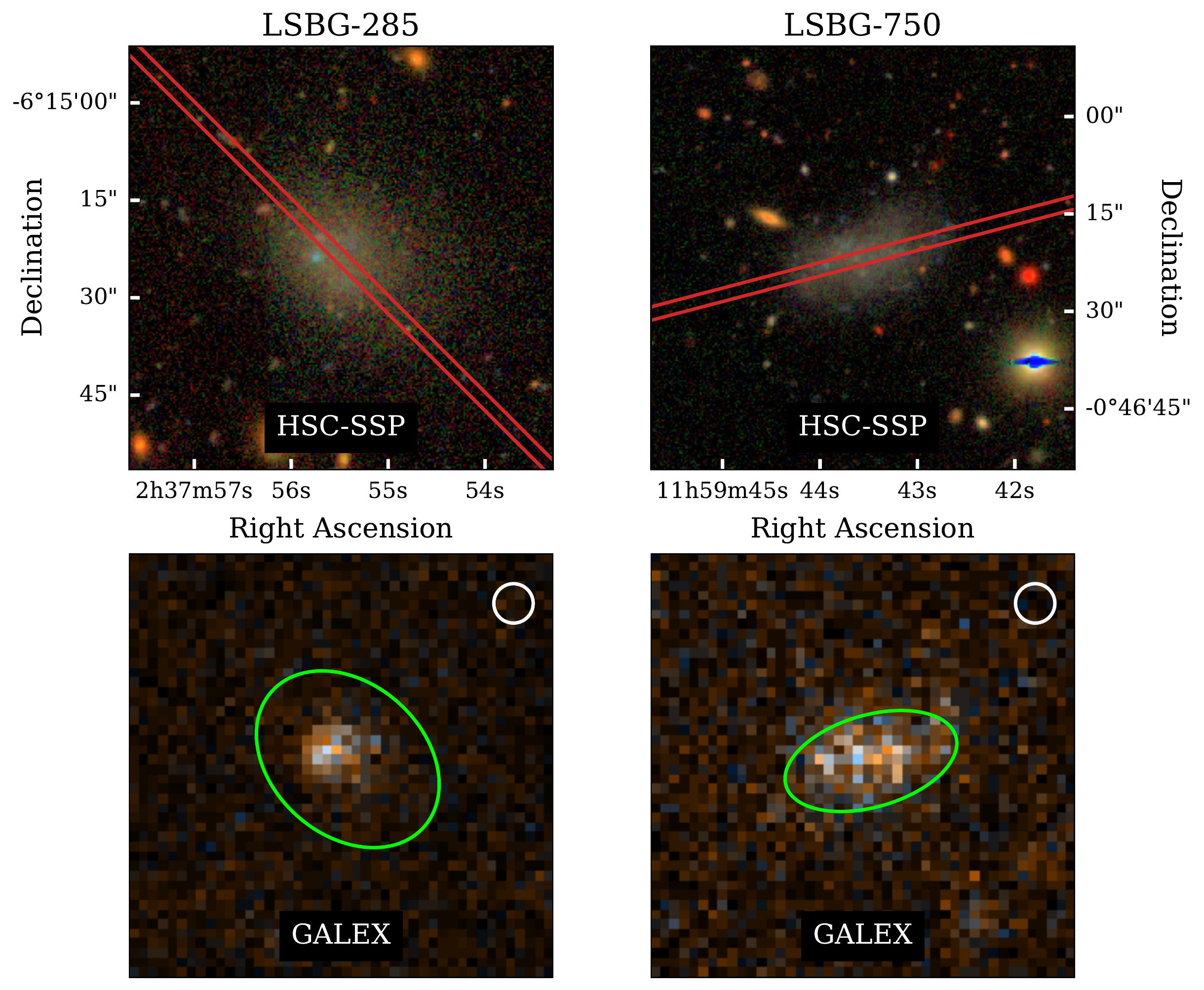}
    \caption{HSC-SSP $gri$ (top row; \citealt{Lupton:2004aa}) and GALEX FUV+NUV (bottom row) composite images  of LSBG-285 (left column) and LSBG-750 (right column). In each HSC-SSP panel, the red lines show the slit position (Section~\ref{sec:obs}). In each GALEX panel, the green ellipse shows the photometric aperture used in our SED fits (Section~\ref{sec:ap-phot}), and the white circle shows the approximate scale of the GALEX PSF. All of the images are 65\arcsec\ on a side.} 
    \label{fig:twoD}
\end{figure*}

\section{Target Selection} \label{sec:target}

The target galaxies were discovered as part of a systematic search for LSB galaxies within the wide layer of the HSC-SSP \citep{Greco:2018ab}. An overview of the HSC-SSP survey design is given in \citet{Aihara:2018aa}, and the first public data release covering ${\sim}100$ deg$^2$ is described in \citet{Aihara:2018ab}. The \citet{Greco:2018ab} galaxy sample contains ${\sim}800$ galaxies, roughly half of which are ultra-LSB, with $g$-band central surface brightnesses $\mu_0(g) > 24$~\sbunit. The sample spans a wide range of galaxy colors and morphologies, with ${\sim}$40\% of sources having blue optical colors ($g-i < 0.7$) and ultraviolet (UV) detections in the Galaxy Evolution Explorer (GALEX) source catalog \citep{Martin:2005aa}. This suggests ongoing star formation in these systems and makes them  promising targets for follow-up emission-line measurements. 

To test the feasibility of obtaining emission-line redshifts of such diffuse targets with a modest amount of telescope time, we selected three of the brightest UV sources from the \citet{Greco:2018ab} sample that were visible from Gemini South during the Gemini Fast Turnaround 2017A semester for follow-up spectroscopic observations. Two of our selected targets were observed (LSBG-285 and LSBG-750), but the third became inaccessible from Gemini South before the observations could be taken. In Figure~\ref{fig:twoD}, we show HSC-SSP $gri$ and GALEX FUV+NUV composite images of our two observed targets (our imaging data are described in Section~\ref{sec:phot-ana}). 

\section{Spectroscopic Data and Analysis}\label{sec:spec-ana}

\begin{figure*}[ht!]
    \centering
    \includegraphics[width=\textwidth]{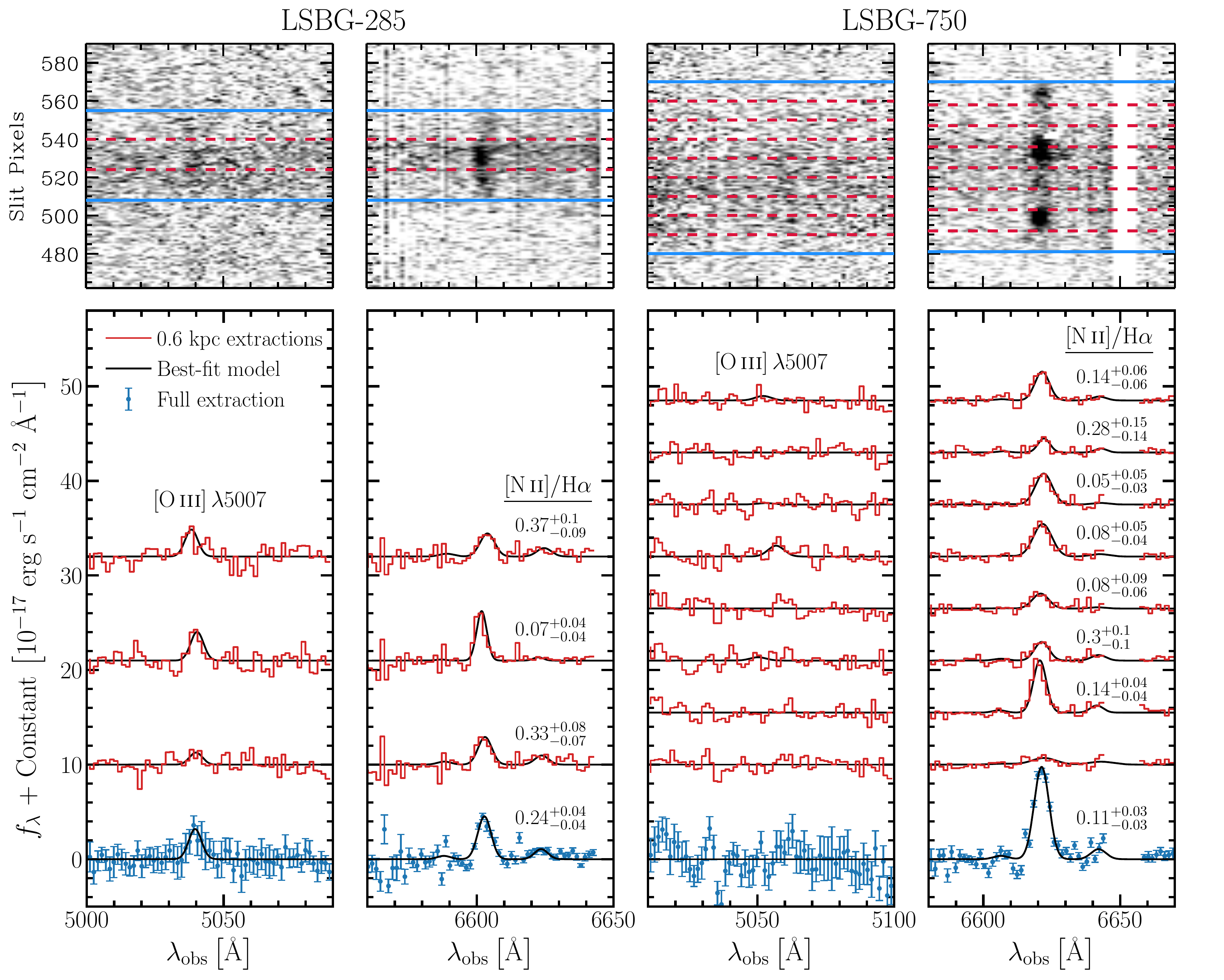}
    \caption{Two- (top) and one-dimensional (bottom) spectra of LSBG-285 (left) and LSBG-750 (right). Red lines show individual 0.6~kpc extractions, and blue points show the full galaxy extraction. The red one-dimensional spectra have been scaled by a factor of two for visibility. The best-fit spectral-line models (Section~\ref{sec:spec-fit}) are indicated by the black lines. The median \NIIb/H$\alpha$ value and associated 16th and 84th percentile uncertainties, which we derive from the marginalized posterior probability distributions, are indicated near each spectrum in the relevant panels.}
    \label{fig:spectra}
\end{figure*}

\subsection{Observations and Data Reduction}\label{sec:obs}

We carried out long-slit spectroscopy of LSBG-285 and LSBG-750 on 2017 July 3 and 13, respectively, using the Gemini Multi-Object Spectrograph (GMOS) on Gemini South under Gemini Science Program GS-2017A-FT-21. Our primary goal was to measure redshifts and line ratios for our targets, rather than to spectrally or spatially resolve the emission-line kinematics. We used the 831 lines mm$^{-1}$ grating with a wide 2\arcsec\ slit to increase our sensitivity for these LSB objects and to increase the probability of including H{\,\sc ii} regions within these galaxies. This configuration produced a spectral resolution of $\sigma\approx3$~\AA, which corresponds to 140~km~s$^{-1}$ at a wavelength of $\lambda=6563$~\AA, \resp{with an observed spectral range of 4880-7200~\AA.}

To increase the per-pixel signal-to-noise ratio, we binned by four pixels in both the spatial and spectral directions, producing a pixel scale of 0$\farcs$3 in the spatial direction and 1.5 \AA\ in the spectral direction. For each source, the slit was centered on the galaxy and oriented along its major axis according to the parameters given in \citet{Greco:2018ab}; the slit positions are overlaid on the HSC-SSP images in Figure~\ref{fig:twoD}. The on-source integration time for each target was 1~hr.

We constructed a master bias by median-combining the evening's calibration bias frames and subtracting the overscan region. Flat fields were constructed by median-combining our observation-specific flat fields, subtracting the master bias, and applying the appropriate GMOS hot pixel mask provided by Gemini. Each target was observed for 20 minutes in 3 exposures. These individual exposures were bias-subtracted and then flat-fielded using our master flat. A gain normalization was applied to the individual CCDs, which were mosaicked into a single image for each exposure. The three exposures were finally combined into an individual image. Cosmic rays were removed using a median filter. We applied the same reduction methodology to the standard star frames. 

We used the arc spectra to apply an approximate wavelength solution to the combined target and standard star mosaicked images, and then used the positions of the night sky lines to further improve the wavelength solution. Since there are few night sky lines in the blue region of our spectra, the precision of the wavelength solution is a function of wavelength, with the highest precision occurring near H$\alpha$. The root-mean-square errors about the wavelength solution were ${\sim}2.0$, 0.4, and 0.5~\AA\ at ${\sim}5000$, 6000, and 7000~\AA, respectively. We subtracted the average background using spectra on either side of the dispersed target spectrum that were free from emission due to sources that serendipitously fell within the slit. The target spectra were then flux calibrated using spectra extracted from the mosaicked standard star. 

We extracted one-dimensional source spectra extending across the full spatial extent of the galaxies within rectangular regions, with lengths of 51 pixels (15\farcs3) and 88 pixels (26\farcs4) for LSBG-285 and LSBG-750, respectively. We further extracted sub-regions in the spatial direction with lengths of 17 pixels ($5\farcs1\approx0.6$~kpc) for LSBG-285 and 11 pixels ($3\farcs3\approx0.6$~kpc) for LSBG-750 to measure changes in the emission features as a function of position. In Figure~\ref{fig:spectra}, we show two- and one-dimensional spectra of each source. 

\subsection{Emission-line Measurements} \label{sec:spec-fit}

Our goal is to recover redshifts and line ratios from the extracted one-dimensional spectra. For each galaxy, we perform the following analysis on the individual 0.6~kpc extractions, as well as the full galaxy extractions, which can be viewed as stacking the individual extractions to produce higher signal-to-noise ratio measurements.

We fit the H$\alpha$ + \NII\ and \OIII\ regions of the each spectrum separately. For the former, we simultaneously fit the data with a flat continuum plus three Gaussian line profiles; since the lines are unresolved, we force the profiles to have the same width. We assume that the H$\alpha$ and [N\,{\sc ii}] emitting regions are at the same redshift, with the redshift measurement being dominated by the much stronger H$\alpha$ emission line. We further force the amplitude of \NIIa\ to be a factor of three below that of \NIIb. Thus, our model of the H$\alpha$ + \NII\ lines has five free parameters: two Gaussian amplitudes, one central wavelength, one profile width, and a constant. When fitting the \OIII\ line†, we assume a flat continuum plus a single Gaussian line profile with standard deviation given by the H$\alpha$ fit, resulting in three free parameters. Note the spectra do not cover [O\,{\sc iii}]$\,\lambda 4959$ or H$\beta$. 

To perform the fits, we assume independent Gaussian uncertainties and construct the standard log-likelihood function $\ln \mathcal{L} \equiv -\chi^2/2$. We first maximize the log-likelihood using \code{L-BFGS-B} optimization implemented in the \code{scipy} software package. The resulting maximum-likelihood parameters are then used as the initialization of a Markov chain Monte Carlo (MCMC) exploration of the posterior probability distribution, where we assume reasonable (e.g., positive amplitudes), uniform proper priors on each parameter. For the MCMC implementation, we use \code{emcee} \citep{emcee}. In Figure~\ref{fig:spectra}, we show our best-fit models overlaid on each spectral extraction from LSBG-285 and LSBG-750.

 The median \NIIb/H$\alpha$ flux ratio and the associated 16th and 84th percentile uncertainties are indicated next to each spectrum in the relevant panel. In some cases, the detection of \NIIb\ is marginal, making the flux ratio highly uncertain. This is particularly true for the LSBG-750 spectra, which have a gap near the \NIIb\ line due to a CCD artifact. Nevertheless, the full extractions, as well as some of the 0.6~kpc extractions, from both galaxies show statistically significant detections of \NIIb. We only detect \OIII\ with a signal-to-noise ratio greater than 2.5 in two out of three of LSBG-285's 0.6~kpc spectral extractions; this line is robustly detected in the full extraction of this galaxy. For all other cases, we calculate an upper limit for the \OIII\ flux as $\sqrt{2\pi}\,\sigma\times$~median(pixel-to-pixel flux error), where $\sigma$ is the width of the H$\alpha$ line. All \OIII\ measurements for LSBG-750 are upper limits.
 
 Within the full galaxy spectral extractions for LSBG-285 and LSBG-750, the integrated H$\alpha$ fluxes are $\log(F_\mathrm{H\alpha}/\mathrm{erg\ cm^{-2}\ s^{-1}}) = -15.56\pm0.03$ and $-15.17\pm0.01$, respectively. For LSBG-285, the integrated \OIII\ flux is $\log( F_\mathrm{O\textsc{iii}}/\mathrm{erg\ cm^{-2}\ s^{-1}}) = -15.7\pm0.1$. These integrated fluxes have not been corrected for Galactic extinction.

\subsection{Redshift and Distance Measurements}\label{sec:distances}

For our redshift measurements, we use the H$\alpha$ emission-line centroid detected in the full galaxy spectral extractions. LSBG-285 and LSBG-750 are diffuse dwarfs galaxies outside of the Local Volume, with redshifts of $cz_\mathrm{helio}=1742\pm19$~km~s${^{-1}}$ and $cz_\mathrm{helio}=2586\pm18$~km~s${^{-1}}$, respectively, where the uncertainties are dominated by our wavelength calibration uncertainty. Given the relatively low recession velocities of these galaxies, we calculate distances using the local velocity field model of \citet{Mould:2000aa}, which accounts for the influence of the Virgo Cluster, the Great Attractor and the Shapley Supercluster. For LSBG-285, we find a velocity correction of $\delta v = -20.5$~km s$^{-1}$, and for LSBG-750, we find $\delta v = 309$~km s$^{-1}$. The derived proper distances are $24.6\pm0.3$ and $41.3\pm0.3$~Mpc, respectively, where the quoted uncertainties only account for our redshift uncertainties.

\section{Photometric Data and Analysis}\label{sec:phot-ana}

\subsection{Datasets and Image Processing}

We study the spectral energy distributions (SEDs) of our sources from the far-UV (FUV) to the mid-infrared (MIR) using archival near-UV (NUV) and FUV imaging from GALEX, $grizy$ imaging from the wide layer of HSC-SSP, and archival W1 (3.4 $\mu$m) imaging from the Wide-field Infrared Survey Explorer (WISE). Here, we describe each dataset and our procedure for obtaining/constructing the associated intensity and variance images, which we use to perform aperture photometry.  

\subsubsection{GALEX} \label{sec:galex}

Both LSBG-285 and LSBG-750 serendipitously fell within archival GALEX pointings. For LSBG-285, we use imaging from GALEX's Nearby Galaxy Survey with an exposure time of 1615~s in both the NUV and FUV filters. The imaging of LSBG-750 was taken as part of a GALEX Guest Investigator Program with an exposure time of 1549.5~s  in both UV filters. For each galaxy, we downloaded the raw count ($C$) and high-resolution relative response ($R$) images from the Mikulski Archive for Space Telescopes GALEX tile retrieval service\footnote{\url{http://galex.stsci.edu}}. The intensity and variance images are then given by 
\begin{equation}
I=C/R\ \ \mathrm{and}\ \ V=C/R^2.
\end{equation}
Note the GALEX intensity images are not background subtracted. We estimate the background in an annulus around each source during the aperture-photometry procedure described below. 

For the GALEX photometry, we estimate the Galactic reddening $E(B-V)$ using the maps of
\citet{Schlegel:1998aa} and the \citet{Cardelli:1989aa} extinction law with
$R_V = A_V/E(B-V) = 3.1$,  $A_\mathrm{FUV}=8.24\,E(B-V)$, and $A_\mathrm{NUV}=8.2\,E(B-V)$ \citep{Wyder:2007aa}.
 
\subsubsection{HSC-SSP}

We use $grizy$ HSC-SSP imaging from the wide survey layer of the S16A internal data release (see \citealt{Aihara:2018ab} for information about HSC-SSP data releases). We directly use the data products produced by the HSC-SSP software pipeline \citep{Bosch:2018aa}. These products include the background-subtracted intensity images, variance images, object masks, and the model point spread function (PSF) at the location of each galaxy.

\subsubsection{unWISE}

We use archival imaging from WISE \citep{WISE:2010} to constrain the MIR flux of the galaxy SEDs, which provides a strong constraint on internal extinction. The ALLWISE Atlas Images were PSF-convolved by the WISE team to produce optimal detection maps, which degraded the native resolution of the images from $\mathrm{FWHM}\sim6\farcs5$ to $\mathrm{FWHM}\sim8\arcsec$. Resolution is important for our aperture photometry measurements, since there is a large uncertainty associated with masking background sources that may dominate the light in a given aperture, particularly given the LSB nature of our sources. Therefore, we use the ``unWISE'' coadds from \citet{Lang:2014aa}, which preserve the resolution of the original WISE images. We downloaded background-subtracted intensity and variance cutout images of each source from the unWISE website\footnote{\url{http://unwise.me}}. 

We inspected both the W1- and W2-band images, and the sources are visible in both bands. However, the detections in W2 are much noisier and likely suffer from over-subtraction (as evidenced by many negative pixels), particularly around LSBG-285. Furthermore, the resolution of W1 is slightly better than that of W2. We therefore choose to exclude W2 from our analysis, noting that including it does not significantly impact any of our results.

\resp{We attempted to use the newest unWISE coadds, which are based on the  Near Earth Object (NEO) WISE Reactivation mission \citep{Meisner:2017aa, Meisner:2017ab}. These coadds reach a depth of coverage ${\sim}$3$\times$ greater than that of the AllWISE Atlas coadds from \citet{Lang:2014aa}, making them potentially better suited for this work. However, we found that our measured uncertainties on the resulting photometry were unreasonably small for both objects, and the W1 flux for both objects was significantly higher than the ALLWISE stacks. This behavior is consistent with previously known sky-subtraction problems with the NEOWISE coadds, and it is recommended to use the original unWISE coadds despite their having larger background noise than the newer ones that incorporate additional NEOWISE data (Meisner, private communication).}

\begin{table}[t!]
\begin{center}
\caption{Matched-aperture Magnitudes} 
\label{tab:mags}
\begin{tabular}{c|cc}
\hline\hline
\multicolumn{1}{c}{Filter}& LSBG-285  & LSBG-750 \\ 
\hline    
NUV   &  $20.66\pm0.09$  & $20.52\pm0.07$\\
FUV	  &	 $20.26\pm0.06$  & $20.31\pm0.05$ \\
$g$	  &  $18.29\pm0.08$  & $19.00\pm0.08$\\
$r$   &  $17.89\pm0.08$  & $18.77\pm0.08$\\
$i$	  &  $17.75\pm0.08$  & $18.69\pm0.08$\\
$z$   &  $17.70\pm0.09$  & $18.62\pm0.09$\\
$y$   &  $17.7\pm0.1$    & $18.6\pm0.1$ \\
W1    &  $19.1\pm0.1$    & $19.9\pm0.2$ \\  
\hline\hline
\end{tabular}
\end{center}
\vspace{-0.3cm}
\tablecomments{All magnitudes are on the AB system and have been corrected for Galactic extinction.}
\end{table}

\subsection{Aperture Photometry} \label{sec:ap-phot}

Our goal is to measure the relative flux in each band within the same physical aperture; we do not attempt to capture all the galaxy light, which would be highly sensitive to the masking of background sources. We use elliptical apertures with parameters based on the catalog of \citet{Greco:2018ab}. To find the aperture size, we determine the radius at which the slope of the growth curve begins to rapidly increase due to background sources entering the aperture. For each object, an aperture with semi-major axis equal to $1.5\times r_\mathrm{eff}$ \resp{provides a good} balance between capturing as much light as possible while avoiding the need to mask background sources; \resp{LSBG-285 has $r_\mathrm{eff}=10\farcs3$ and LSBG-750 has $r_\mathrm{eff} = 9\farcs0$.} These apertures are shown on the GALEX images in Figure~\ref{fig:twoD}. We note that the optical colors measured within these apertures are consistent at the ${\sim}0.01$~mag level with the colors based on the total model magnitudes measured by \citet{Greco:2018ab}.

\begin{figure*}[t!]
    \centering
    \includegraphics[width=\textwidth]{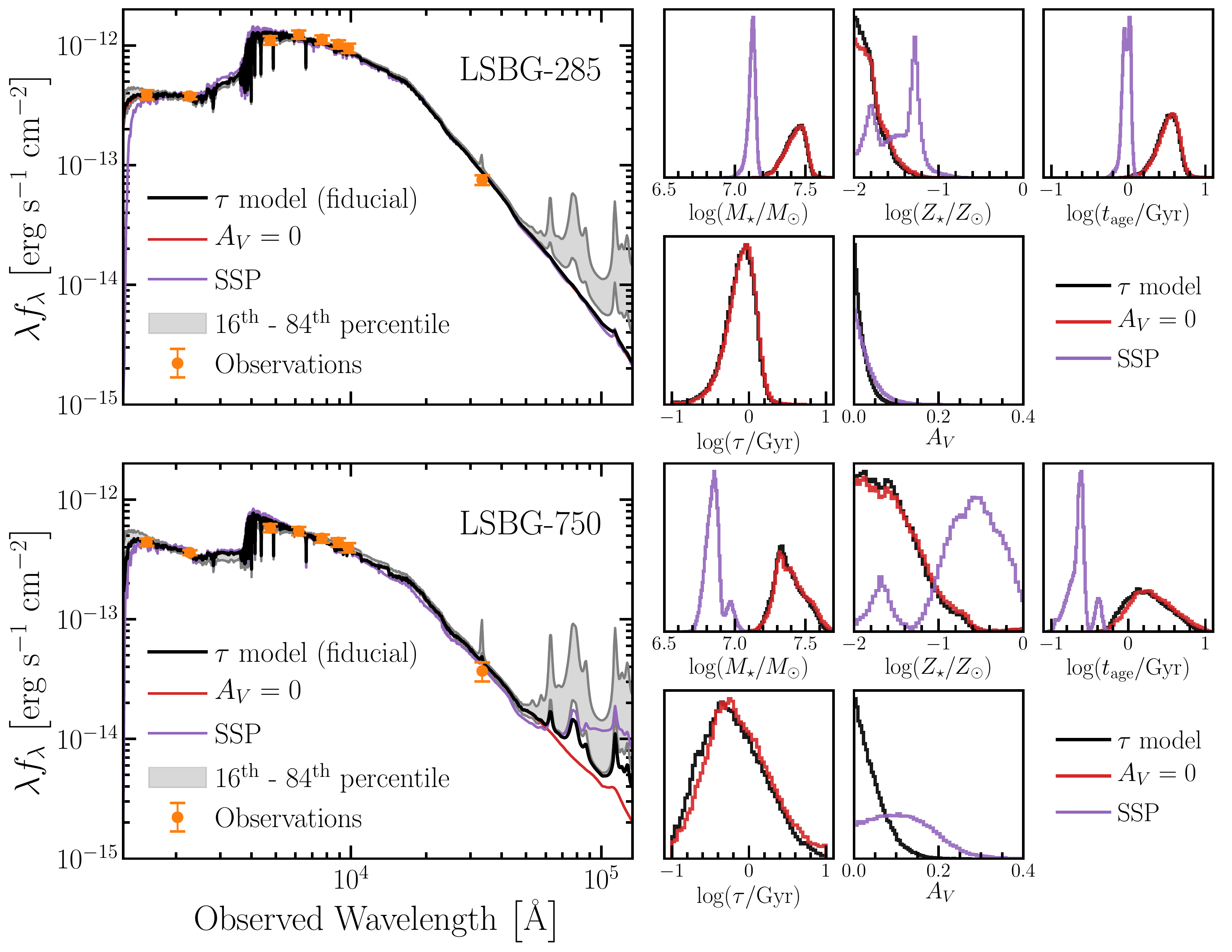}
    \caption{{\it Left:} Observed photometry and the maximum a posteriori model spectra assuming a $\tau$-model star formation history with dust (our fiducial model), without dust ($A_V=0$), and a simple stellar population with dust (SSP; $\tau=0$). The gray shaded regions show the 16th-84th percentile $\tau$-model fluxes in each wavelength bin. {\it Right:} Marginalized parameter distributions for each of the model assumptions shown in the left panel. We assume uniform priors for each parameter, with the bounds given in Section~\ref{sec:sed-fit}. The 50th percentile parameter values and the associated 16th and 84th percentile uncertainties are given in Table~\ref{tab:properties}.}
    \label{fig:SED}
\end{figure*}

To ensure we are measuring light from the same physical components of the galaxies, we match the resolution of each image to that of the WISE W1 band, which has the broadest PSF (FWHM $\sim6\farcs5$). For the unWISE and GALEX images, we transform the PSFs to a Gaussian of width FWHM $\sim6\farcs5$ using the convolution kernels provided by \citet{Aniano:2011aa}. For the HSC-SSP images, we assume Gaussian PSFs and smooth each image with a Gaussian with standard deviation 
\begin{equation}
\sigma = \sqrt{\sigma^2_\mathrm{target} - \sigma_\mathrm{intr}^2},
\end{equation}
where $\sigma_\mathrm{intr}$ is the intrinsic resolution as measured by fitting a Gaussian to the model PSF at the location of each galaxy, and the target resolution is given by $2\sqrt{2 \ln 2}\, \sigma_\mathrm{target} = 6\farcs5$. To propagate the uncertainties from the image convolutions, we convolve each variance image with the square of the kernel used on the associated intensity image. 

The photometric measurements are then given by 
\begin{align}
F &= \sum_i \tilde{I}_i,\\
\sigma^2_F &= \sum_i \tilde{V}_i,
\end{align}
where the sum is over the pixels in the aperture, $F$ is the measured flux, $\sigma_F$ is the associated error estimate, and the tildes indicate that these are the images after the convolutions described above. We carry out these photometric measurements using the Astropy affiliated package \code{photutils} \citep{photutils}.

As noted in Section~\ref{sec:galex}, the GALEX images have not been background-subtracted. We therefore subtract a background level estimated within an annulus of inner radius $4\,r_\mathrm{eff}$ and outer radius $7\,r_\mathrm{eff}$ for each galaxy, with the appropriate error propagation term added to the variance. For the HSC-SSP photometry, \citet{Greco:2018ab} estimate a typical uncertainty of 0.08~mag associated with sky subtraction near these sources, which we add our HSC-SSP error estimates. We tabulate our matched-aperture photometric measurements in Table~\ref{tab:mags}. 

\subsection{SED Fitting} \label{sec:sed-fit}

We study the FUV -- MIR SED of each galaxy using the aperture photometry from Section~\ref{sec:ap-phot} and the Bayesian inference code \code{prospector}\footnote{\url{https://github.com/bd-j/prospector}} \citep{prospector}. This software generates model SEDs on the fly using the Flexible Stellar Population Synthesis package \citep[\code{FSPS};][]{Conroy:2009aa} and explores the potentially high-dimensional posterior probability distribution via MCMC sampling. We run \code{FSPS} using the \citet{Calzetti:2000} extinction curve, the MILES spectral templates \citep{MILES1, MILES2}, and the Padova isochrones \citep{Padova1, Padova2}. We implement MCMC using \code{emcee}. See \citet{Leja:2017aa} and \citet{Pandya:2018aa} for other applications of \code{prospector}.

Most of the parameters of \code{FSPS} may be free parameters in \code{prospector}. Our fiducial model assumes an exponentially declining star-formation history ($\tau$ model). We fix the source redshifts at their observed values and allow five parameters to float: stellar mass $M_\star$, stellar metallicity $Z_\star$, age since the first onset of star formation $t_\mathrm{age}$, e-folding timescale $\tau$, and internal extinction as parametrized by the $V$-band extinction coefficient $A_V$. All other parameters are fixed at their default values\footnote{The default parameters are listed at \url{http://dfm.io/python-fsps}.}. For comparison purposes, we also run the analysis with a simple stellar population (SSP; $\tau=0$~Gyr), as well as a dust-free model ($A_V = 0$). We assume the following uniform priors: $M_\star = 10^6$ to $10^{10}~M_\odot$ , $[Z_\star/Z_\odot] = -2.0$ to $0.2$, $\log_{10}(\tau/\mathrm{Gyr}) = -1.0$ to 1.0, $\log_{10}(\mathrm{t_\mathrm{age}/Gyr}) = -3.0$ to 1.15, and $A_V = 0$ to 4. We sample logarithmically in $\tau$ and $t_\mathrm{age}$ and linearly in all other parameters. 

\resp{Hydrodynamic cosmological simulations suggest that extended dwarf galaxies may have undergone bursty star formation histories \citep[e.g.,][]{DiCintio:2017aa, Chan:2018aa}. However, it is difficult to observationally distinguish a smoothly declining star formation history from a bursty, episodic one \citep{Ruiz-Lara:2018aa}. This is particularly true for a photometric SED analysis such as we present here. As a test, we ran our full analysis assuming a single burst scenario combined with a $\tau$ model, and the inferred parameters are very similar to our simpler fiducial model.}

In Figure~\ref{fig:SED}, we show our photometric measurements and the maximum a posteriori model spectra from our various \code{prospector} runs. We also show the marginalized parameter distributions for each case. Both SEDs are consistent with very little to no dust, regardless of whether we assume our fiducial $\tau$-model star formation history or an SSP, although the SSP does tend to predict higher $A_V$ values (particularly for LSBG-750) to compensate for the lack of an older population of red stars. The marginalized parameter distributions for our models with and without dust are therefore very similar.

Focusing on our fiducial model, the stellar populations of both LSBG-285 and LSBG-750 are consistent with very low stellar metallicity ([$Z_\star/Z_\odot] < -1.0$) and intermediate age ($t_\mathrm{age} \approx 3$~Gyr and 2~Gyr), with extended star formation histories characterized by low e-folding timescales ($\tau \approx 0.8$~Gyr and 0.6~Gyr). The $\tau$ posterior distribution for both galaxies falls to very low values before it reaches $\tau=0.1$~Gyr (the boundary of our prior), suggesting that an exponentially declining star formation history is preferred over an SSP. 

Compared to our fiducial $\tau$ model, the SSP produces significantly different marginalized posterior distributions for the stellar mass, stellar metallicity, and age of the system. The SSP generally predicts stellar masses that are ${\sim}0.3$-$0.5$~dex lower and ages that are ${\sim}0.6$-$0.8$~dex lower than the $\tau$ model. Furthermore, the marginalized distributions for the stellar metallicities become bimodal in the SSP case, with the metallicity posterior of LSBG-750 reaching to super-solar values. 

The SSP results demonstrate the behavior of the marginalized posterior distributions in the limit that $\tau\rightarrow0$~Gyr. However, it is reasonable to assume that multiple generations of stars exist within these galaxies. In this scenario, the integrated light is easily dominated by the youngest stellar population, while the mass is dominated by somewhat older stars. For the remainder of the paper, we therefore assume the results from our fiducial $\tau$ model, which are consistent with low, but nonzero, e-folding timescales $\tau$. 

\section{Galaxy Properties}\label{sec:galprops}

We now combine results from our spectroscopic (Section~\ref{sec:spec-ana}) and photometric (Section~\ref{sec:phot-ana}) analyses to study the environments and physical properties of LSBG-285 and LSBG-750. A summary of the galaxy properties is provided in Table~\ref{tab:properties}. 

\subsection{Environments} \label{sec:redshift-env}

To investigate the galaxy environments, we search for neighbors using the NASA-Sloan Atlas\footnote{\url{http://nsatlas.org}} (NSA), which contains virtually all galaxies with known redshifts out to $z = 0.055$ within the coverage of Sloan Digital Sky Survey (SDSS) DR8 \citep{Aihara:2011aa}. The NSA also provides stellar mass estimates calculated using \code{kcorrect} \citep{Blanton:2007aa}, which assumes the initial mass function of \citet{Chabrier:2003aa} and is based on fits to both the SDSS optical and, when available, GALEX fluxes.

Both LSBG-285 and LSBG-750 are quite isolated. Based on the NSA, LSBG-285's nearest neighbor is a dwarf galaxy with $M_\star = 4\times10^7~M_\odot$ at a comoving separation of 1.2~Mpc, and the nearest galaxy with $M_\star > 10^{10}~M_\odot$ is at a distance of 3.4~Mpc. Similarly, LSBGs-750's nearest neighbor is a $M_\star = 2\times10^7~M_\odot$ dwarf at a comoving separation of 1.6~Mpc, and the nearest galaxy with $M_\star > 10^{10}~M_\odot$ is 4.5~Mpc away. For context, \citet{Geha:2012aa} found that essentially all galaxies with $M_\star \lesssim 10^8~M_\odot$ that are separated by more than 1.5~Mpc from a massive host are star-forming, where massive hosts are defined to have $M_\star\sim2.5\times10^{10}~M_\odot$. In other words, low-mass quenched galaxies only exist within a few virial radii of massive hosts. These authors define galaxies beyond 1.5~Mpc from a massive host to be in the field. The star-forming nature and isolation of our galaxies are consistent with this picture. \resp{While both galaxies are in field-like environments, LSBG-750 is in a much lower density environment, with two galaxies with $M_\star > 10^{7}~M_\odot$ within 3~Mpc, whereas LSBG-285 has eight such dwarf galaxy neighbors (according to the NSA galaxy catalog).}

The NSA catalog follows the footprint of SDSS DR8 and thus does not cover the entire sky. LSBG-285 and LSBG-750 fall 3.5$^\circ$ (1.5~Mpc) and 2.8$^\circ$ (2~Mpc) from the SDSS footprint boundary, respectively. Therefore, the most conservative statement about their environments is that they are at least these distances from a massive host. In the case of LSBG-285, the distribution of NSA galaxies around its location reflect the pattern of SDSS stripes, suggesting varying levels of completeness in this region. We searched the NASA/IPAC Extragalactic Database for neighbors around LSBG-285 that might be missing from the NSA galaxy catalog; no additional massive neighbors were found.

\resp{Our environment measures are prone to error due to large relative velocities between the potential hosts and the target galaxies. Thus, as an additional measure of the isolation of these galaxies, we calculate their projected separations from all NSA galaxies with  $M_\star = 2.5\times10^{10}~M_\odot$ within a redshift range of $|\Delta z|\cdot c = 500$ km~s$^{-1}$. In this case, the nearest massive neighbors to LSBG-285 and LSBG-750 are at projected distances of 0.9~Mpc and 4.7~Mpc, respectively. Using this isolation metric, both galaxies are in low-density, field-like environments, and LSBG-750 in particular appears to be extremely isolated.} 

\begin{table}[t!]
\caption{Galaxy Properties} 
\label{tab:properties}
\begin{tabular}{lr|rr}
\hline\hline
\multicolumn{2}{l}{Observed Property}  & LSBG-285  & LSBG-750 \\
\hline
R.A. &  J2000   & 02$^\m{h}$37$^\m{m}$55$^\m{s}$\!.48 & 11$^\m{h}$59$^\m{m}$43$^\m{s}$\!.55 \\
Decl. &  J2000  & $-06^\circ$15$^\prime$23\farcs86   & $-00^\circ$46$^\prime$21\farcs76 \\
$cz_\mathrm{helio}$            & km s$^{-1}$           &  $1742\pm19$ & $2586\pm18$  \\
$m_g$             & mag  &  $18.0\pm0.2$ & $18.6\pm0.2$ \\
$\mu_0(g)$        & $\m{mag\ arcsec^{-2}}$  &  $24.1\pm0.4$ & $24.2\pm0.4$ \\
$g-r$             &  mag &  $0.4\pm0.1$      &  $0.2\pm0.1$ \\
$g-i$             &  mag &   $0.5\pm0.1$         &  $0.3\pm0.1$ \\
$\mathrm{NUV}-r$             &  mag &   $2.37\pm0.09$         &  $1.54\pm0.09$ \\
$r_\mathrm{eff}$  & arcsec  & $10.3\pm0.8$ & $9.0\pm0.8$ \\
Ellipticity   \hspace{-.2cm}     &        &  $0.25\pm0.03$ &  $0.48\pm0.03$ \\
S\'{e}rsic $n$    &        &  $0.7\pm0.3$ &  $0.6\pm0.3$ \\
\hline
\multicolumn{2}{l}{Derived Property}  & LSBG-285  & LSBG-750 \\ 
\hline
Distance\footnote{Distances assume the flow model of \citet{Mould:2000aa}. Quoted uncertainties only account for our redshift uncertainties.}     & Mpc	   &  $24.6\pm0.3$	 & $41.3\pm0.3$ \\
$M_g$             &  mag &  $-14.0\pm0.2$  & $-14.5\pm0.3$ \\
$r_\mathrm{eff}$  & kpc    &  $1.2\pm0.1$ &  $1.8\pm0.2$ \\
$M_\star$         & $10^7\ M_\odot$ &  $2.7^{+0.4}_{-0.5}$ & $2.3^{+0.9}_{-0.6}$ \\ 
$A_V$             &    mag   &  ${<}\,0.06$ & ${<}\,0.11$\\ 
$[Z_\star/Z_\odot]$\footnote{Stellar metallicity.}  &        & ${<}-1.5$ & ${<}-1.0$ \\ 
$t_\mathrm{age}$\footnote{Age since the first onset of star formation.}  & Gyr             &  $3.3^{+1.0}_{-0.9}$ & $1.7^{+1.7}_{-0.8}$ \\ 
$\tau$\footnote{Star formation history e-folding timescale.}         & Gyr             &  $0.8^{+0.4}_{-0.3}$ & $0.6^{+1.0}_{-0.3}$ \\ 
12 + log(O/H)\footnote{Median oxygen abundance of spectral extractions, where the error is the spread in values. The Sun has 12 + log(O/H)= 8.66.}  \hspace{-2cm}  &        & $8.4\pm0.2$ & $8.4\pm0.3$ \\
\vspace{-0.35cm}\\
\hline\hline
\end{tabular}
\tablecomments{Coordinates, total magnitudes, surface brightnesses, and structural measurements are from \citet{Greco:2018ab}. All magnitudes are on the AB system, and they have been corrected for Galactic extinction using the dust map of \citet{Schlegel:1998aa} and the recalibration from \citet{Schlafly:2011aa}. Parameters inferred from our SED fits assume an exponentially declining star formation history, and we quote the median, 16th, and 84th percentile of the marginalized posterior distributions. Upper limits are the 95th percentile of the posterior distributions. For distance-dependent quantities, we conservatively add in quadrature a systematic uncertainty equal to the difference between our adopted parameter value and the value derived from assuming pure Hubble flow.}
\end{table}

\subsection{Stellar Masses and Effective Radii}

The photometric apertures we used for SED fitting only contain a fraction of the total galaxy light. We therefore assume that the inferred stellar mass-to-light ratios are uniform with radius and use the total $i$-band magnitudes from \citet{Greco:2018ab} to estimate the total stellar masses; these magnitudes are based on two-dimensional S\'{e}rsic function fits to the HSC-SSP images. We find that LSBG-285 has $M_\star=2.7\times10^7~M_\odot$ and LSBG-750 has $M_\star=2.3\times10^7~M_\odot$. To estimate the physical extent of these galaxies, we again use the single S\'{e}rsic function fits from \citet{Greco:2018ab}, finding that LSBG-285 and LSBG-750 have effective radii of $r_\mathrm{eff} = 1.2$ and 1.8~kpc, respectively. Hence, both are consistent with being ${\sim}10^7~M_\odot$ LSB dwarf galaxies at the small-size end of the UDG population. 

These results are summarized in Table~\ref{tab:properties} as the 50th, 16th, and 84th percentiles of the marginalized posterior distributions shown in Figure~\ref{fig:SED}. Since these galaxies are relatively nearby, their distances have added uncertainty due to the local velocity field. For all distance-dependent quantities in  Table~\ref{tab:properties}, we adopt the values inferred from assuming the \citet{Mould:2000aa} flow model. For each of these  parameters, we conservatively add in quadrature a systematic uncertainty equal to the difference between our adopted parameter value and the value derived from assuming pure Hubble flow. 

 \begin{figure*}[t!]
    \centering    
	\includegraphics[width=\textwidth]{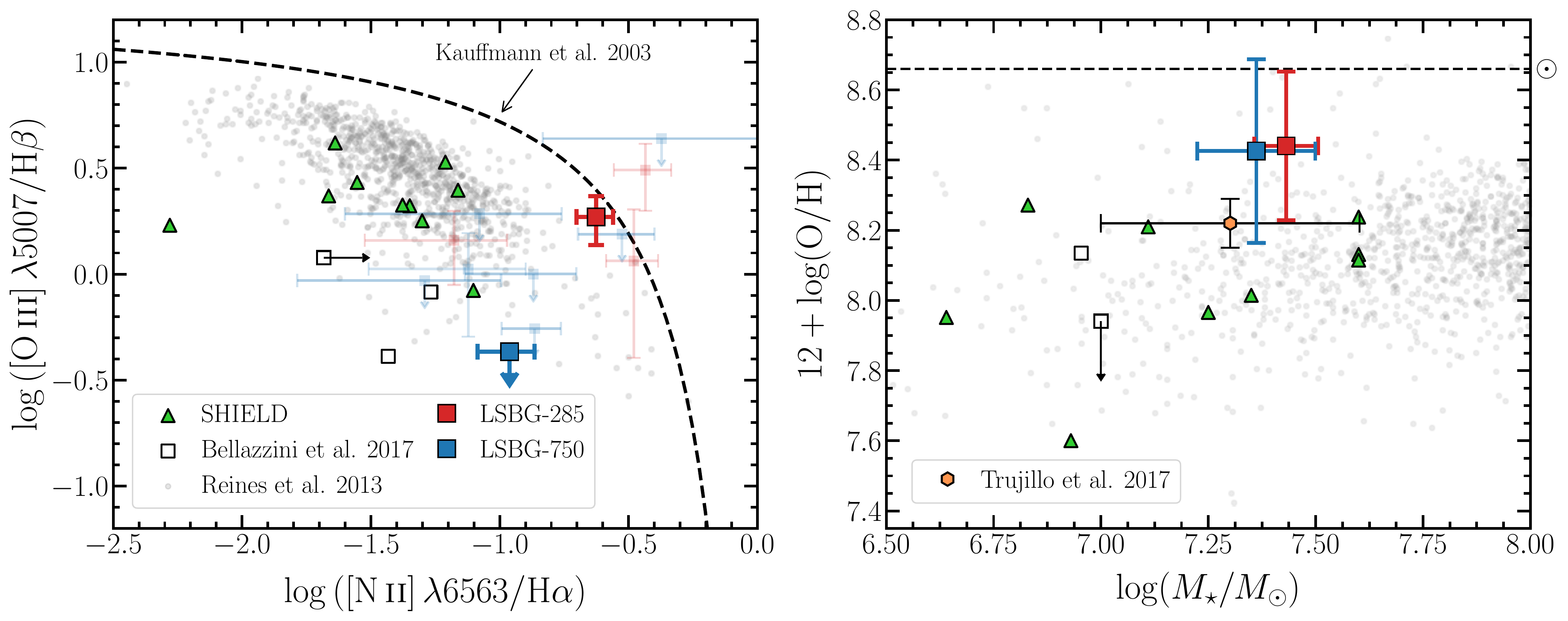}
    \vspace{-0.5cm}
    \caption{{\it Left:} Emission-line diagnostic diagram based on our full galaxy spectral extractions (large filled squares) and individual 0.6~kpc extractions (small transparent squares). Upper/lower limits are indicated by arrows. The dashed curve divides the diagram into regions that are typically occupied by star-forming nebulae and active galactic nuclei \citep{Kauffmann:2003aa}. We also show measurements from SHIELD galaxies \citep{Haurberg:2015aa}, two extended dwarf irregulars \citep{Bellazzini:2017aa}, and low-mass star-forming galaxies selected from the NSA by \citet{Reines:2013aa}. {\it Right}: Oxygen abundance--stellar mass relation for the same samples as the left panel. \resp{The error bars for LSBG-285 and LSBG-750 show the spread in metallicity between the individual spectral extractions. In addition, the nearby star-forming UDG UGC 2162 \citep{Trujillo:2017aa} is indicated by the orange hexagon.} Oxygen abundances were derived from the linear relation of \citet{Pettini:2004aa}. Solar abundance is indicated by the dashed black line.}
    \label{fig:BPT}
\end{figure*}

\subsection{Star Formation Rates}

Assuming the distances we derived from the flow model of \citet{Mould:2000aa}, we calculate star formation rates (SFRs) using the measured FUV luminosities and the scaling relation from \citet{Kennicutt:1998ab}:
\begin{equation}\label{eqn:sfr_fuv}
  \mathrm{SFR}\ [M_\odot\ \mathrm{yr^{-1}}] = \frac{L_\nu(\mathrm{FUV})}{7.14\times10^{27}\ \mathrm{erg\ s^{-1}\ Hz^{-1}}},
  \end{equation}
where $L_\nu(\mathrm{FUV})$ is the Galactic extinction-corrected FUV luminosity measured within an elliptical aperture with semi-major axis equal to $2.5\times r_\mathrm{eff}$ for LSBG-285 and $4\times r_\mathrm{eff}$ for LSBG-750. The inferred SFRs are ${\sim}0.002$ and $0.009~M_\odot$ yr$^{-1}$, respectively, where we have not corrected for internal extinction from dust. As our SED fits show (Section~\ref{sec:sed-fit}), both galaxies likely have $A_V<0.1$~mag. If we assume the 95th percentile upper limits on $A_V$ given in Table~\ref{tab:properties} and the extinction law of \citet{Cardelli:1989aa} with $R_V = A_V/E(B-V) = 3.1$, the FUV luminosity correction leads to SFRs that are higher by ${\sim}10$\% for LSBG-285 and ${\sim}20$\% for LSBG-750. In either case, these SFRs are low compared to gas-rich galaxies of similar stellar mass detected by ALFALFA \citep{Huang:2012ab}, though they are consistent with the large scatter observed in this relation.  

For normal spiral galaxies with $\mathrm{SFR}\sim1~M_\odot$ yr$^{-1}$, FUV-derived SFRs largely agree with those estimated from H$\alpha$ nebular emission after accounting for internal dust attenuation; however, for dwarf galaxies with $\mathrm{SFR}<0.1~M_\odot$ yr$^{-1}$, H$\alpha$ SFRs have been observed to systematically underpredict the total SFR relative to FUV SFRs \citep{Lee:2009aa}. To test if our sources follow this trend, we estimate the H$\alpha$ SFR as \citep{Kennicutt:1998ab}: 
\begin{equation}\label{eqn:sfr_ha}
  \mathrm{SFR}\ [M_\odot\ \mathrm{yr^{-1}}] = \frac{L(\mathrm{H}\alpha)}{1.26\times10^{41}\ \mathrm{erg\ s^{-1}}},
  \end{equation}
where $L(\mathrm{H}\alpha)$ is the Galactic extinction-corrected integrated H$\alpha$ luminosity. To compare the FUV and H$\alpha$ SFRs, we recalculate the FUV SFR using our smaller SED apertures and apply a rough aperture correction to $L(\mathrm{H}\alpha)$ given by the ratio of the areas of the photometric and spectral apertures. This assumes that the H$\alpha$ emission extends throughout the photometric apertures, which, given the patchiness seen in the two-dimensional spectra, likely overestimates $L(\mathrm{H}\alpha)$ (particularly in the case of LSBG-285).

For LSBG-285, H$\alpha$ predicts a higher SFR than FUV, with $\mathrm{SFR(H\alpha)/SFR(FUV)\sim1.5}$. In contrast, for LSBG-750, the two SFRs are consistent, with $\mathrm{SFR(H\alpha)/SFR(FUV)\sim1}$. Both of these values are consistent with the scatter observed in this ratio for galaxies of similar stellar mass and SFR \citep{Lee:2009aa}. Note that these are very approximate calculations, particularly given the very uncertain aperture corrections for H$\alpha$. 

\subsection{Gas-phase Metallicities}

We robustly detect H$\alpha$ in both LSBG-285 and LSBG-750. As described in Section~\ref{sec:spec-fit}, we also have marginal detections of \NIIb\ for both sources, suggesting it may be possible to use the \NIIb/H$\alpha$ flux ratio as a probe of gas-phase metallicity. However, we must first verify that the observed emission is consistent with being due to H{\sc\, ii} regions. For this purpose, we use the emission-line diagnostic diagram of \citet[][the BPT diagram]{Baldwin:1981aa}, \NIIb/H$\alpha$ vs. \OIII/H$\beta$. While we have upper limits and/or detections of \OIII\ in both objects, H$\beta$ is outside the wavelength range of our observations. Nevertheless, our SED fitting results suggest there is very little dust in these galaxies ($A_V<0.1$), making it reasonable to assume H$\alpha$/H$\beta$ = 2.86. This is the intrinsic ratio corresponding to a temperature of 10$^4$~K and electron density of 10$^2$~cm$^{-3}$ for Case B recombination \citep{Osterbrock:1989aa}. 

Assuming this value for H$\alpha$/H$\beta$, we show our sources on the BPT diagram in the left panel of Figure~\ref{fig:BPT}. Our full galaxy spectral extractions have the highest signal-to-noise ratio and are indicated by the red (LSBG-285) and blue (LSBG-750) filled squares. Our measurements within 0.6~kpc apertures are indicated by small transparent squares with error bars. The dashed curve divides this plane into regions expected to be occupied by star-forming nebulae and active galactic nuclei \citep{Kauffmann:2003aa}. Within the uncertainties, all of our measurements confirm that LSBG-285 and LSBG-750 fall within the star-forming locus of the BPT diagram, as expected. 

The gray points in Figure~\ref{fig:BPT} show ${\sim}1000$ low-mass galaxies ($M_\star\sim 10^{6.5}$-$10^8~M_\odot$) selected from the NSA galaxy catalog by \citet{Reines:2013aa}, who were searching for active galactic nuclei in dwarf galaxies. The green triangles show measurements of H{\,\sc ii} regions in \HI-selected dwarfs from the SHIELD project \citep{Cannon:2011aa, Haurberg:2015aa}. With \HI\ masses as low as ${\sim}10^{6.5}~M_\odot$, the SHIELD galaxies were selected to sample the very-low-mass end of the \HI\ mass function. We also show measurements from H{\,\sc ii} regions in two dwarf irregular galaxies (open squares) with UDG-like sizes and surface brightnesses \citep{Bellazzini:2017aa}.

The measurements from our sources tend to have higher \NIIb/H$\alpha$ and lower \OIII/H$\beta$ than most of the systems in Figure~\ref{fig:BPT}, albeit with large uncertainties in both flux ratios. This trend in the BPT diagram---moving down and to the right---is known to correlate with increasing gas-phase metallicity \citep[e.g.,][]{Moustakas:2006aa}. In the right panel of Figure~\ref{fig:BPT}, we show oxygen abundance as a function of stellar mass for the same systems shown in the left panel. For each measurement, we calculate the oxygen abundance assuming the linear relation from \citet{Pettini:2004aa}:
\begin{equation}\label{eqn:N2-to-OH}
	12 + \log(\mathrm{O/H}) = 8.9 + 0.57\,\mathrm{N2},
\end{equation}
where $\mathrm{N2}\equiv$~log(\NIIb/H$\alpha$). Both LSBG-285 and LSBG-750 are consistent with having high gas-phase metallicity for their stellar mass. \resp{The nearby star-forming UDG UGC~2162 also has a relatively high gas-phase metallicity for its stellar mass of ${\sim}2\times10^7~M_\odot$ \citep{Trujillo:2017aa}; this object is indicated by the orange hexagon in the right panel of Figure~\ref{fig:BPT}.} As we obtain more optical spectra of blue UDGs, both optically and \HI\ selected, it will be interesting to see if this trend continues.

In is important to note that the above results are tentative given the relatively low signal-to-noise ratio of our measurements and the uncertainties associated with applying Equation~(\ref{eqn:N2-to-OH}) to such diffuse systems. This relation was calibrated with samples of individual H{\sc ii} regions with direct-method metallicity measurements; however, diffuse ionized gas not contained in H{\sc ii} regions also contributes significantly to the optical line emission of galaxies, and the addition of this component may lead to elevated \NIIb/H$\alpha$ compared to single H{\sc ii} regions \citep{Sanders:2017aa}, which would bias our metallicity estimate high. Finally, we note that galaxy samples (particularly optically selected samples) in this mass and surface brightness range are highly incomplete, so it is currently difficult to assess the significance of the above comparison. 

\subsection{Ordered Rotation in LSBG-750}\label{sec:rotation}

\begin{figure}[t!]
    \centering
    \includegraphics[width=\columnwidth]{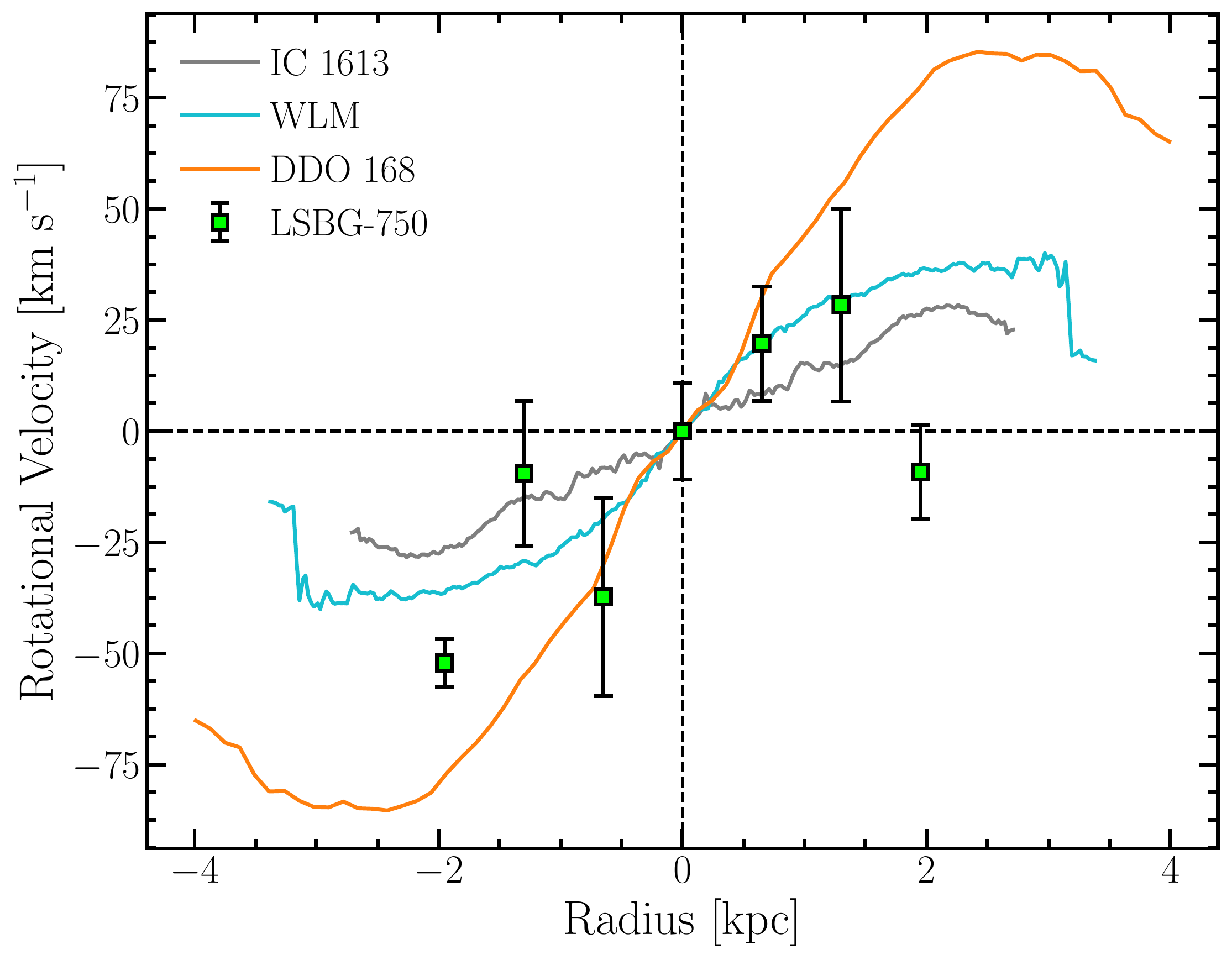}
    \caption{Evidence for ordered rotation in LSBG-750 with a maximum rotational velocity of $\lesssim50$~km s$^{-1}$. As a qualitative comparison, we show high-resolution \HI\ rotation curves of dwarf galaxies with stellar masses of ${\sim}10^7$-$10^8~M_\odot$, which have estimated halo masses ranging from 10$^9~M_\odot$ (IC~1613) to 10$^{11}~M_\odot$ (DDO 168; \citealt{Oh:2015aa}). The velocities have been corrected for inclination.}
    \label{fig:rotation}
\end{figure}

As can be seen by careful inspection of Figure~\ref{fig:spectra}, the wavelength offsets of LSBG-750's individual spectral extractions show some evidence of ordered rotation, whereas LSBG-285's H$\alpha$-emitting region is more concentrated (allowing 3 versus 8 extractions on the same physical scale) with offsets that are not consistent with rotation based on the current data. Our current data lack the signal-to-noise ratio needed to perform detailed mass modeling based on LSBG-750's observed rotation curve. Nevertheless, in Figure~\ref{fig:rotation}, we make a qualitative comparison with high-resolution (${\sim}6\arcsec$ angular and $<2.6$~km s$^{-1}$ velocity resolution) \HI\ rotation curves \citep{Oh:2015aa} of dwarf galaxies of similar stellar mass (${\sim}10^7$-$10^8~M_\odot$) from the LITTLE THINGS survey \citep{Hunter:2012aa}. For LSBG-750, we assume the center of the H$\alpha$-emitting region as the central position and velocity. 

All the velocities have been corrected for inclination, where for LSBG-750, we use the ellipticity measured from the $i$-band image and assume it is an oblate spheroid with an edge-on axis ratio of 0.2 \citep[e.g.,][]{Holmberg:1958aa}. We note that the ellipticity distribution of our full LSB galaxy catalog suggests that on average objects \resp{in our sample} are intrinsically round \citep{Greco:2018ab}. However, the above inclination correction will only increase the velocities, which for our purposes is the more conservative assumption.   

For comparison, the dwarf galaxies IC~1613, WLM, and DDO~168 (Figure~\ref{fig:rotation}) have measured dynamical masses of $\log(M_\mathrm{dyn}/M_\odot)\sim8.4$, 9.0, and 9.5, respectively \citep{Oh:2015aa}. Based on the mass models of \citet{Oh:2015aa}, their estimated halo masses are  $\log(M_\mathrm{halo}/M_\odot)\sim9$, 10, 11, respectively. The H$\alpha$-emitting region in LSBG-750 roughly extends across the entire optical galaxy. The \HI\ may extend beyond the H$\alpha$ in this galaxy, but there is often good agreement (at the level of our uncertainties) between H$\alpha$ and \HI\ maximum rotational velocities in LSB galaxies \citep[e.g.,][]{deBlok:2002aa, Marchesini:2002aa}. We thus infer that LSBG-750 has a maximum rotational velocity $\lesssim50$~km s$^{-1}$, which corresponds to a dynamical mass of $\lesssim10^9~M_\odot$ at ${\sim}2$~kpc. At $M_\star\sim10^7~M_\odot$, a rotational velocity of ${\sim}50$~km s$^{-1}$ is relatively high, but it is consistent with the large scatter observed in the stellar mass Tully-Fisher relation \citep[e.g.,][]{Torres-Flores:2011aa, Bradford:2016aa}. Based on this upper limit and the qualitative comparison with similar objects in Figure~\ref{fig:rotation}, LSBG-750 likely occupies a dwarf-mass dark matter halo with $M_\mathrm{halo} < 10^{11}~M_\odot$. \resp{This result adds to the growing list of rotational measurements of blue \citep{Trujillo:2017aa, Leisman:2017aa, Spekkens:2018aa} and red \citep{Ruiz-Lara:2018aa} UDGs that are consistent with dwarf-like halo masses.}

\section{Discussion}\label{sec:discussion}

The \citet{Greco:2018ab} galaxy sample---from which LSBG-285 and LSBG-750 were selected---contains ${\sim}800$ LSB galaxies, and a significant fraction of these galaxies are blue and UV emitting. The present work is a pilot study for a more systematic follow-up effort to map out the spatial distribution and physical properties of this larger galaxy sample. The optical and environmentally-blind selection of our diffuse-galaxy sample nicely complements most previous work, which generally covers small volumes \citep[e.g.,][]{Dalcanton:1997aa} and/or are biased by either environment \citep[e.g.,][]{van-Dokkum:2015aa} or gas fraction \citep[e.g.,][]{Leisman:2017aa}. Here, we place the first two sources we have followed up in context with UDGs and the general dwarf galaxy population. 

\begin{figure}[t!]
    \centering
    \includegraphics[width=\columnwidth]{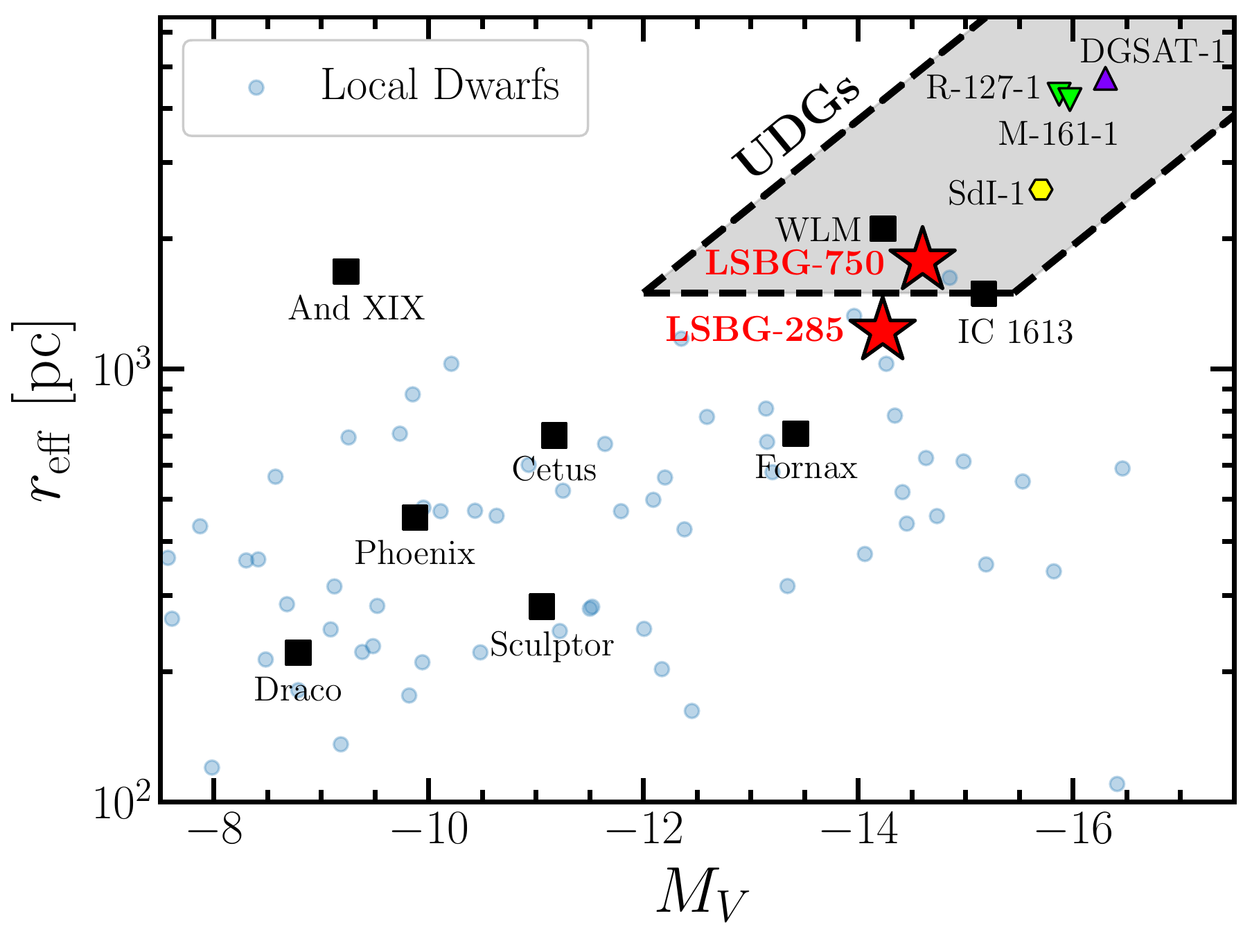}
    \caption{Size-luminosity relation for our two sources (red stars), dwarf galaxies in and around the Local Group \citep[filled circles;][]{McConnachie:2012aa}, and ultra-diffuse galaxies (UDGs; gray shaded region; \citealt{van-Dokkum:2015aa}) in low-density environments \citep{Dalcanton:1997aa, Martinez-Delgado:2016aa, Bellazzini:2017aa}. The labeled black squares show well-known dwarf galaxies from the Local Group.}
    \label{fig:mag-vs-size}
\end{figure}

 In Figure~\ref{fig:mag-vs-size}, we show LSBG-285 and LSBG-750 in the size--luminosity plane, along with dwarf galaxies in and around the Local Group \citep{McConnachie:2012aa}. We have labeled well-known local dwarfs such as Sculptor (the prototypical dwarf spheroidal galaxy), Fornax, and the Wolf-Lundmark-Melotte (WLM) irregular galaxy. The shaded gray region shows roughly where UDGs fall within this parameter space \citep{van-Dokkum:2015aa}. We also show other known isolated UDGs such as the \HI-rich SECCO-dI-1 \citep{Bellazzini:2017aa} and the intriguingly quiescent R-127-1 and M-161-1 \citep{Dalcanton:1997aa}. \resp{SECCO-dI-2 \citep{Bellazzini:2017aa} is another example of an \HI-rich, relatively isolated UDG, but it has similar size and luminosity to our sources and is not shown for clarity. DGSAT-1 \citep{Martinez-Delgado:2016aa} has a red color and exists in a filamentary region near the Pisces-Perseus supercluster with at least two $M_\star>10^{10}~M_\odot$ neighbors within 1~Mpc \citep{Papastergis:2017aa}.} Where necessary, we convert $gr$ measurements to $V$-band using the transformation $V=g-0.59\,(g-r)-0.01$ \citep{Jester:2005aa}. 
 
We note that our full sample is not expected to contain many (if any) objects in the Local Volume (distances within ${\sim}$10~Mpc), since such nearby sources have very large sizes on the sky\footnote{Our search was carried out on $12^\prime\times12^\prime$ patches with 17\arcsec\ overlapping regions, limiting our sensitivity to objects with large angular diameters.} (see \citealt{Danieli:2018aa} for a study of the discovery space for integrated-light searches for LSB dwarfs within the Local Volume). Nonetheless, we are finding field galaxies at the high end of the Local Group luminosity and size distributions. Andromeda~XIX (And~XIX), an M31 satellite, is an interesting object, as its large size for its luminosity is likely due to tidal interactions with its massive host \citep{Collins:2013aa}. WLM and IC~1613 are two other well-known Local Volume objects with UDG-like luminosities and sizes; we compare the rotation curves of these sources with that of LSBG-750 in Figure~\ref{fig:rotation}. \resp{It is interesting to note that, in contrast to our sources and the isolated UDGs mentioned above, Local Volume UDG candidates identified in the Updated Nearby Galaxy Catalog \citep{Karachentsev:2013aa} appear to be associated exclusively with massive neighbors \citep{Karachentsev:2017aa}.}

\resp{For UDGs that exist in galaxy clusters or groups \citep[e.g.,][]{Yagi:2016aa, van-der-Burg:2017aa}, environmental processes such as ram pressure stripping may be responsible for their ultra-low stellar densities \citep[e.g.,][]{Yozin:2015aa}. In contrast, for isolated galaxies such as LSBG-285 and LSBG-750 \citep[see also, for example,][]{Dalcanton:1997aa, Bellazzini:2017aa}, internal processes are required to explain their structural properties.} In an \HI\ follow-up study of four relatively isolated UDGs, \citet{Papastergis:2017aa} found that they fell into two categories---one is \HI\ rich and star-forming and the other is apparently gas-poor and quiescent. 

Our sources likely fit into the former category. If they follow the stellar density--color--gas fraction relation from \citet{Huang:2012ab}, they are expected to have gas fractions of $M_\mathrm{HI}/M_\star\sim1.3-3.0$. If this \HI\ content is \resp{found}, their extended star formation histories may be consistent with the formation scenario proposed by \citet{DiCintio:2017aa}, in which feedback-induced outflows lead to the expansion of the dark matter and stellar distributions, \resp{and for isolated systems, a significant \HI\ gas mass is predicted}. 

It is also possible that our sources represent the high-spin tail of the dwarf galaxy population \citep{Amorisco:2016aa}, which in the case of LSBG-750, is also consistent with our data (Section~\ref{sec:rotation}); higher signal-to-noise ratio optical spectra and/or \HI\ kinematic measurements are necessary to properly test this scenario.   

Similar to the few UDGs whose stellar populations have been studied spectroscopically \citep{Kadowaki:2017aa, Gu:2017aa, Ruiz-Lara:2018aa, Ferre-Mateu:2018aa} and photometrically \citep{Pandya:2018aa}, both our sources are consistent with the observed stellar mass--metallicity relation for dwarf galaxies \citep{Kirby:2013aa}, which is roughly continuous with the relation for galaxies as massive as $M_\star=10^{12}~M_\odot$. \resp{However, these previously studied UDGs are all quenched systems}---LSBG-750 extends this relation to star-forming UDGs in the field with stellar masses in the ${\sim}10^7~M_\odot$ range. \resp{Whether or not there is an evolutionary link between blue and red UDGs remains an open question, though there are hints that such a link exists \citep[e.g.,][]{Roman:2017ab}}.

The similar stellar mass--metallicity relation for cluster and field UDGs is at least consistent with these objects having similar formation histories, with differences due to environment occurring at late times. Furthermore, this supports the hypothesis that most UDGs are extensions of the general dwarf galaxy population. However, it is \resp{interesting} that the gas-phase metallicities appear to be relatively high in our two sources \resp{and UGC 2162 (see Figure~\ref{fig:BPT}), all of which have relatively large physical sizes for their stellar mass. In particular, the opposite trend is observed in galaxies of higher stellar mass ($M_\star \gtrsim 5\times10^8~M_\odot$), where smaller sizes at fixed stellar mass correspond to higher gas-phase metallicities \citep[e.g.,][]{Ellison:2008aa, Snchez-Almeida:2018aa}. While it is possible that the elevated metallicities in these diffuse systems are due to the deep gravitational potential wells of over-massive dark matter halos, as has been observed in a Coma UDG (\citealt{van-Dokkum:2016aa}; although, see \citealt{DiCintio:2017aa}), the current rotational measurements of both LSBG-750 and UGC 2162 \citep{Trujillo:2017aa} point towards typical stellar-to-halo mass ratios for these galaxies. We again caution that there are large uncertainties associated with the \NIIb/H$\alpha$--oxygen abundance relation \citep{Sanders:2017aa}, which we have used to infer the gas-phase metallicities.} 

Based on the redshifts we currently have in hand (both archival redshifts and those presented here) for galaxies in the \citet{Greco:2018ab} sample, we are finding diffuse galaxies at distances in the range ${\sim}20$-200~Mpc. Our follow-up program will therefore significantly expand the volume out to which it is possible to study optically-selected, LSB dwarf galaxies. This effort, combined with complementary \HI\ \citep[e.g.,][]{Giovanelli:2013aa, Tollerud:2015aa} and optical \citep[e.g.,][]{Danieli:2018aa} searches for LSB dwarfs within the Local Volume, will provide a much more complete picture of the low-luminosity, LSB dwarf galaxy population, which will have important implications for our understanding of galaxy formation within the \LCDM\ cosmological framework.

\section{Summary}\label{sec:summary}

We have presented a follow-up study of two diffuse dwarf galaxies in the field, LSBG-285 and LSBG-750, which were recently discovered with the HSC-SSP \citep{Greco:2018ab}. These galaxies live outside the Local Volume at comoving distances of ${\approx}25$ and ${\approx}41$~Mpc, respectively. There are no massive galaxies ($M_\star > 10^{10}~M_\odot$) within at least 1.5~Mpc from LSBG-285 and 2~Mpc from LSBG-750. Both objects are physically large compared to most dwarfs in and around the Local Group (Figure~\ref{fig:mag-vs-size}); LSBG-285 has $r_\mathrm{eff}=1.2$~kpc and LSBG-750 has $r_\mathrm{eff}=1.8$~kpc, making them similar in size and surface brightness to ultra-diffuse galaxies (UDGs). However, they are distinct from most known UDGs in that they are star-forming, exist in the field, and were selected in the optical in an environmentally blind survey. In the case of LSBG-750, we set an upper limit on its rotational velocity of ${\lesssim}50$~km~s$^{-1}$, which is comparable to dwarf galaxies of similar stellar mass with estimated halo masses of $<10^{11}~M_\odot$ (Figure~\ref{fig:rotation}). We summarize the observed and inferred galaxy properties in Table~\ref{tab:properties}.

We studied the stellar populations of these systems using UV--MIR matched-aperture photometry and the Bayesian SED fitting code \code{prospector}, assuming an exponentially declining star formation history for our fiducial model (see Figure~\ref{fig:SED} for the marginalized posterior distributions). The stellar populations of both objects are likely of intermediate age (${\sim}1$-3~Gyr) and have undergone extended star formation histories characterized by low, but nonzero, e-folding timescales ($\tau<1$~Gyr). Their current star formation rates (${\sim}0.002$-0.01~$M_\odot$ yr$^{-1}$) are low compared to gas-rich galaxies of similar stellar mass detected by ALFALFA, though they are consistent with the large scatter observed in \resp{the star formation rate-stellar mass relation for such objects.} With stellar metallicities $[Z_\star/Z_\odot] \lesssim -1.0$ and total stellar masses ${\sim}10^7~M_\odot$, both galaxies are consistent with the observed stellar mass--metallicity relation for dwarf galaxies.

Based on our measurements of (or limits on) \OIII, \NIIb, and H$\alpha$ combined with the very low dust content in these galaxies (as evidenced by our SED fits), LSBG-285 and LSBG-750 fall on the star-forming locus of the BPT diagram, suggesting that we are observing emission from H{\,\sc ii} regions within these diffuse systems. Assuming the linear relation from \citet{Pettini:2004aa} to convert \NIIb/H$\alpha$ into oxygen abundance, our sources are consistent with having somewhat high gas-phase metallicity for their stellar mass (Figure~\ref{fig:BPT}). Higher signal-to-noise ratio observations and a better understanding of the true distribution of galaxies at such low surface brightnesses will be required to place these results in context with the better-studied, higher-surface-brightness galaxy population.   

\software{
  This work additionally utilized \code{astropy}
  \citep{Astropy-Collaboration:2013aa}, \code{numpy}
  \citep{Van-der-Walt:2011aa}, \code{scipy} (\url{https://www.scipy.org}),
  \code{matplotlib} \citep{Hunter:2007aa}, \code{sfdmap} 
  (\url{https://github.com/kbarbary/sfdmap}), and \code{corner} \citep{corner:2016}.
}

\acknowledgments

\resp{We thank the anonymous referee for their careful reading of our paper and 
their useful comments and suggestions.}
We thank Gonzalo Aniano for making his convolution kernels publicly 
available and for his assistance with generating additional WISE kernels. We 
thank Amy Reines for sharing her dwarf galaxy catalog. J.P.G.
is grateful to Adrian Price-Whelan and Viraj Pandya for useful conversations.
J.P.G. was supported by the National Science Foundation under grant 
No. AST 1713828.

The Hyper Suprime-Cam (HSC) collaboration includes the astronomical communities
of Japan and Taiwan, and Princeton University. The HSC instrumentation and
software were developed by the National Astronomical Observatory of Japan
(NAOJ), the Kavli Institute for the Physics and Mathematics of the Universe
(Kavli IPMU), the University of Tokyo, the High Energy Accelerator Research
Organization (KEK), the Academia Sinica Institute for Astronomy and
Astrophysics in Taiwan (ASIAA), and Princeton University. Funding was
contributed by the FIRST program from Japanese Cabinet Office, the Ministry of
Education, Culture, Sports, Science and Technology (MEXT), the Japan Society
for the Promotion of Science (JSPS), Japan Science and Technology Agency (JST),
the Toray Science Foundation, NAOJ, Kavli IPMU, KEK, ASIAA, and Princeton
University. 

This paper makes use of software developed for the Large Synoptic Survey
Telescope. We thank the LSST Project for making their code available as free
software at http://dm.lsst.org.

Based in part on data collected at the Subaru Telescope and retrieved from the
HSC data archive system, which is operated by Subaru Telescope and Astronomy
Data Center, National Astronomical Observatory of Japan.

The spectra presented in this work are based on observations obtained at the Gemini Observatory, which is operated by the Association of Universities for Research in Astronomy, Inc., under a cooperative agreement with the NSF on behalf of the Gemini partnership: the National Science Foundation (United States), the National Research Council (Canada), CONICYT (Chile), Ministerio de Ciencia, Tecnolog\'{i}a e Innovaci\'{o}n Productiva (Argentina), and Minist\'{e}rio da Ci\^{e}ncia, Tecnologia e Inova\c{c}\~{a}o (Brazil).

This research has made use of the NASA/IPAC Extragalactic Database (NED) which is operated by the Jet Propulsion Laboratory, California Institute of Technology, under contract with the National Aeronautics and Space Administration.

\bibliography{ref}

\end{document}